# Unsupervised Domain Adaptation for MRI Volume Segmentation and Classification Using Image-to-Image Translation

Satoshi Kondo[1] and Satoshi Kasai[2]

[1] Muroran Institute of Technology, Hokkaido, Japan
[2] Niigata University of Health and Welfare, Niigata, Japan

**Abstract.** Unsupervised domain adaptation is a type of domain adaptation and exploits labeled data from the source domain and unlabeled data from the target one. In the Cross-Modality Domain Adaptation for Medical Image Segmentation challenge (crossMoDA2022), contrast enhanced T1 MRI volumes for brain are provided as the source domain data, and high-resolution T2 MRI volumes are provided as the target domain data. The crossMoDA2022 challenge contains two tasks, segmentation of vestibular schwannoma (VS) and cochlea, and classification of VS with Koos grade. In this report, we presented our solution for the crossMoDA2022 challenge. We employ an image-to-image translation method for unsupervised domain adaptation and residual U-Net the segmentation task. We use SVM for the classification task. The experimental results show that the mean DSC and ASSD are 0.614 and 2.936 for the segmentation task and MA-MAE is 0.84 for the classification task.

## 1  Introduction

Unsupervised domain adaptation (UDA) is a type of domain adaptation and exploits labeled data from the source domain and unlabeled data from the target one [1, 2]. In the Cross-Modality Domain Adaptation for Medical Image Segmentation challenge held in MICCAI2022 conference (crossMoDA2022), a large and multi-class dataset for unsupervised domain adaptation is introduced [3]. In this challenge, contrast enhanced T1 MRI volumes for brain are provided as the source domain data, and high-resolution T2 MRI volumes are provided as the target domain data [4, 5]. The crossMoDA2022 challenge contains two tasks as followings.

a) Task 1 - Segmentation of two key brain structures (tumor and cochlea) involved in the follow-up and treatment planning of vestibular schwannoma (VS). The diagnosis and surveillance in patients with VS are commonly performed using contrast-enhanced T1 (ceT1) MR imaging. However, there is growing interest in using non-contrast imaging sequences such as high-resolution T2 (hrT2) imaging due to improvement of patient safety and cost efficacy.

b) Task 2 - Classification of VS according to the Koos grade in hrT2 images. The Koos grading scale is a classification system for VS that characterizes the tumor and its impact on adjacent brain structures. There are four grades. Grade 1 means small intracanalicular tumor, grade 2 means small tumor with protrusion into the cerebellopontine cistern and no contact with the brainstem, grade 3 means tumor occupying the cerebellopontine cistern with no brainstem displacement, and grade 4 means large tumor with brainstem and cranial nerve displacement. Koos grading is currently performed on ceT1 scans, but hrT2 could be used.

In this report, we present our solution to the crossMoDA2022 challenge. We employ an image-to-image translation method for unsupervised domain adaptation and residual U-Net with deep super vision for the segmentation task. We use support vector machines (SVM) with hand-crafted features for the classification task.

## 2    Proposed Method

The dataset includes contrast enhanced T1 MRI (ceT1) volumes for brain as the source domain data (the segmentation and the classification labels are provided), and high-resolution T2 MRI (hrT2) volumes as the target domain data without any labels. In the proposed method, ceT1 volumes are translated to hrT2-like volumes by using an image-to-image translation method. The segmentation of VS and cochlea is performed with our segmentation model which is trained by using the translated hrT2-like volumes and the segmentation labels for corresponding ceT1 volumes. Classification of VS according to the Koos grade is conducted by using SVM with hand-crafted features obtained from the segmentation results. We will explain the details of the image-to-image translation method, the segmentation model and the classification model in the followings.

We use DCLGAN [6] as our image-to-image translation method. DCLGAN is an unsupervised image-to-image translation method. It is based on contrastive learning and a dual learning setting (exploiting two encoders) to infer an efficient mapping between unpaired data. We apply DCLGAN to translate ceT1 slices to hrT2 slices, i.e. the translation is conducted in not 3D volumes but 2D images.

We use 3D encoder-decoder networks for the segmentation task. Our base model is residual U-Net with deep super vision [7]. The input volumes for the training is translated volumes from ceT1 to hrT2 with DCLGAN. An input volume is resampled in [0.4 mm, 0.4 mm, 0.5 mm] for x, y and z direction, respectively, at first. MRI volumes are normalized with clipping. The minimum and maximum values are 26 and 486, respectively, for the clipping. In the training phase, we randomly sample 3D patches from the input volumes. The size of a 3D patch is 96 x 96 x 96 voxels. The ratio of positive, i.e., VS and cochlea, and negative patches in the sampling for one input volume is 1:1:1. We apply intensity shift within 5 % for augmentation.

The loss function is adaptive t-vMF Dice loss [9]. The parameter $\lambda$ for the adaptive t-vMF Dice loss is set to 256. We also employ deep super vision for loss calculation. Intermediate outputs from several layers in the decoder of the model are up-sampled, loss value is calculated for each up-sampled output, and then the loss values are aggregated. The number of layers used in the deep super vision is three.

We train multiple models. Each model is trained independently using different combinations of training and validate datasets, and the inference results are obtained by ensemble of the outputs from the models. The final likelihood score is obtained by averaging the likelihood scores from the models. We use five models in our experiments.

We conduct the classification of VS according to the Koos grade by using SVM with hand-crafted features. The features used for the classification are the volume of VS and the size of the bounding box of VS in x, y and z directions, where we use the segmentation results of VS for calculating these features. We use linear-type SVM and the SVM is trained by using the segmentation label of VS for ceT1 volumes.

## 3    Experiments

Our method is implemented by mainly using PyTorch [10], PyTorch Lightning and MONAI libraries. We use three Nvidia RTX3090 GPUs for training.

The crossMoDA2022 dataset contains 210 ceT1 volumes (source) with segmentation and classification labels and 210 hrT2 volumes (target) without labels for training, and 64 hrT2 volumes for validation.

For the training of DCLGAN, we randomly selected about 4,000 slices from ceT1 volumes and hrT2 volumes, respectively. The optimizer for the training of DCLGAN is Adam [7] and the learning rate changes with cosine annealing. The initial learning rate is 0.0002. The number of epoch is 200. The model at the last epoch is selected as the final model. Figure 1 shows an example of ceT1 to hrT2 translation with DCLGAN.

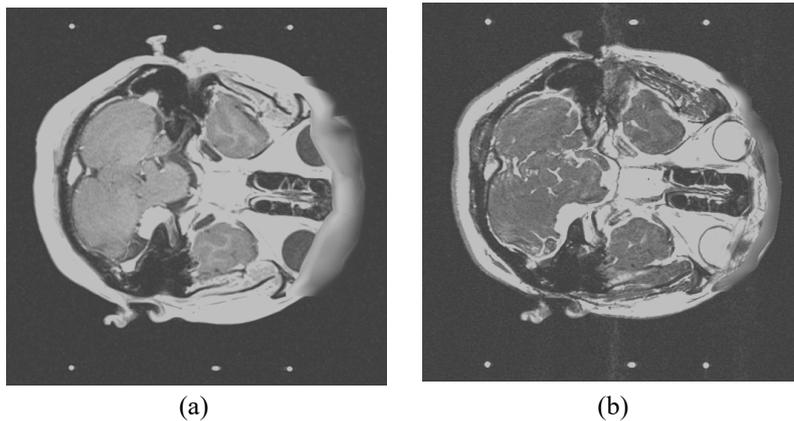

(a)                                    (b)

Figure 1. An example of image-to-image translation results. (a) Input ceT1 slice.
(b) Translated hrT2-like slice with DCLGAN.

As for the hyper-parameter tuning in DCLGAN, we changed the number of slices from about 500 to about 12,000 for each modality. The learning rates were changed from 1e-5 to 1e-2.

For the training of the segmentation model, the optimizer is Adam and the learning rate changes with cosine annealing. The initial learning rate is 0.001. The number of epoch is 300. The model taking the lowest loss value for the validation dataset is selected as the final model.

As for the hyper-parameter tuning in the segmentation model, the learning rates were change from 1e-5 to 1e-2. We also tried different loss function such as Dice + cross entropy loss.

We evaluated our method with the evaluation system provided by the organizers of crossMoDA2022. For task 1 (segmentation), the Dice Score (DSC) and the Average Symmetric Surface Distance (ASSD) are used as evaluation metrics. For task 2 (classification), macro-averaged mean absolute error (MA-MAE) is used as evaluation metrics.

The results of our submission in task 1 are the mean/std DSC values are $0.450 \pm 0.286$ and $0.779 \pm 0.051$ for VS and cochlea, respectively. And the mean/std ASSD values are $5.61 \pm 8.21$ and $0.264 \pm 0.156$ for VS and cochlea, respectively. The result of our submission in task 2 is that MA-MAE is 0.84.

## 4      Conclusions

In this report, we presented our solution for the crossMoDA2022 challenge. We employ an image-to-image translation method for unsupervised domain adaptation and residual U-Net the segmentation task. We use SVM for the classification task. The experimental results show that the mean DSC and ASSD are 0.614 and 2.936 for the segmentation task and MA-MAE is 0.84 for the classification task.